\documentstyle[aps,prb,special,epsf]{revtex}

\newcommand{\V}[1]{\bf #1}                  % Vektor \V{a}% 
\begin{document}
\draft{
  \title{Symmetry breaking in the Hubbard model at weak coupling}
  \author{T.\ Schauerte and P.G.J.\ van Dongen}
  \address{Institut f\"ur Physik, Universit\"{a}t Mainz, 55099, Germany}

  \date{\small \today}

  \maketitle

  \begin{abstract}
    {\rm The phase diagram of the Hubbard model is studied at weak
      coupling in two and three spatial dimensions. It is shown that
      the N\'eel temperature and the order parameter in $d=3$ are
      smaller than the Hartree-Fock predictions by a factor of $q =
      0.2599$.  For $d=2$ we show that the self-consistent (sc)
      perturbation series bears no relevance to the behavior of the
      exact solution of the Hubbard model in the symmetry-broken
      phase. We also investigate an anisotropic model and show that
      the coupling between planes is essential for the validity of
      mean-field-type order parameters.}
  \end{abstract}
  
}

\narrowtext 

The Hubbard model\cite{Hub63} is one of the most important and most
prominent theoretical models in modern condensed matter physics.
Originally introduced in order to describe magnetism in transition
metals, the Hubbard model has been most intensively investigated since
Anderson's proposal that the model should capture the essential
physics of cuprate superconductors.\cite{And87} The
Hubbard-Hamiltonian
\begin{equation}
\label{Hubbard}
{\cal H} = -t \sum_{\langle \V{i}\V{j} \rangle \sigma} 
c^{\dag}_{\V{i}\sigma} c^{\phantom{\dag}}_{\V{j}\sigma} 
+  U \sum_{\V{i}} n_{\V{i}\uparrow} n_{\V{i}\downarrow}
\end{equation}
describes itinerant electrons with spin $\sigma$ on a lattice with
nearest neighbor sites $\langle {\bf i j} \rangle$, interacting
through short-ranged Coulomb repulsion $U$. The success of this
relatively simple lattice model for mobile interacting electrons is
based on its ability to explain a number of important phenomena in
condensed matter physics.  Among these are the (Mott-Hubbard)
metal-insulator transition,\cite{MBR,Geb98} 
antiferromagnetism,\cite{And63} ferromagnetism,\cite{Wah98} 
incommensurate phases,\cite{Schulz90} phase separation,\cite{Vis74,Don95} 
and normal-state properties of high-$T_c$ materials.\cite{And87,Lee92}

In spite of the prominence of the Hubbard model in condensed matter
theory and the deceptively simple two-parameter form of its
Hamiltonian, comparatively little is known exactly (or even
accurately) about its solution.

%%%% d=1 %%%%

In $d=1$ the exact solution has been
determined by the Bethe Ansatz technique.\cite{Lie68}
Even in the extreme weak coupling regime the
ground state of the one-dimensional Hubbard model is
nonperturbative, possessing a discontinuity at $U=0^+$
where the Mott-Hubbard gap opens.

%%%% d=infty %%%%

Much has recently been learned about the solution in high spatial
dimensions ($d=\infty$), but even here\cite{MV,MH89,Georges96} the
exact analytical solution is unknown, so that one has to take recourse
to perturbative\cite{Don95,Don91,Halvorsen94,Georges92} or 
numerical\cite{Georges96,Jarrell92} methods. There is no doubt that the
half-filled Hubbard model exhibits long-ranged antiferromagnetism in
the high-dimensional limit, whereas the situation away from
half-filling is less clear.

%%%% d=2 %%%%

In $d=2$ much is known from exact diagonalization\cite{Dagotto94} or
Monte Carlo simulation of finite systems.\cite{Hirsch85} In addition,
much effort is spent on the analytical\cite{Schulz90}
and numerical\cite{Schweitzer91,HF-FLEX,DCA,RG} analysis of perturbative and
nonperturbative results. While there seems to be now a consensus that
at half-filling the ground state has long-range antiferromagnetic
order, there are still controversies when the system is doped and
-- even at half-filling -- concerning the question of whether
there is or is not a precursor pseudogap.

%%%% d=3 %%%%

In so far as numerical work on the {\em three-dimensional\/} Hubbard
model has been carried out,\cite{Hirsch87,Shiba87,Scalettar,Staudt}
the extrapolation of the results to the thermodynamic limit is clearly
made difficult by the small linear system size. Much
effort has been invested in the analysis of the phase diagram using
{\em perturbative\/} techniques. In the {\em strong coupling\/} regime
($U\to\infty$), where the half-filled Hubbard model reduces to an
effective Heisenberg model, high-temperature series expansions and
$1/S$-expansions can reliably be used to estimate the
N\'eel temperature and the ground state energy, respectively.
However the analysis of the {\em intermediate\/} and {\em weak
  coupling\/} regime is much more difficult: For this regime a variety
of approaches has been proposed, e.g., by Kakehashi,\cite{Kakehashi88}
Logan,\cite{Logan95} Cyrot, \cite{Cyrot96} and
Dar{\'e}.\cite{Dare00} These approaches all yield estimates for the
N\'eel temperature which reduce to the Hartree-Fock result at weak
coupling. However, it is well-known from studies of the Hubbard model,
based on the $1/d$-expansion\cite{Don91} or the local
approximation,\cite{Niki} that the Hartree-Fock approximation {\em
  overestimates\/} N\'eel temperature and order parameter in $d=3$
by a factor {\em of the order of four}, even in the extreme
weak-coupling limit. The precise value of this renormalization factor
in $d=3$ is as yet unknown.

%%%% outline %%%%

The goal of this paper is, {\em first}, to present exact results
for the broken-symmetry phase of the three-dimensional Hubbard model
at weak coupling. In particular, we present asymptotically exact formulas
for the N\'eel temperature and order parameter. It will be seen that this
asymptotic formula also yields a useful approximation formula for the
N\'eel temperature for all $U \alt D$, where $D=12t$ is the band width
of a three-dimensional hypercubic lattice. {\em Second}, we address
sc perturbation theory for the broken-symmetry phase in
low dimensions ($d=2$) in the ground state. Our main result here is
that sc perturbation theory breaks down altogether, and that low-order
sc perturbation theory bears {\em no\/} relevance to the behavior of the
exact solution of the Hubbard model in $d=2$.  We conclude that the
antiferromagnetic order parameter cannot have a mean-field form. This
result has obvious relevance for theories of
high-$T_{c}$ superconductivity, many of which are based on
perturbation theory in strictly two-dimensional systems. {\em Third},
we investigate the order parameter of the anisotropic Hubbard model,
where two-dimensional planes are weakly coupled in $c$-direction by a
small hopping amplitude $t_\perp \ll t_\parallel$.  We demonstrate
that the sc perturbation series converges -- however sluggishly -- as long
as the interplane hopping $t_\perp$ is finite.

%%%% formalism %%%%

Since the behavior of the Hubbard model even in the weak-coupling
limit is {\em nonperturbative\/} one has to apply self-consistent
theories. There are several ways of imposing
self-consistency.\cite{Baym61,Bickers89,Yedidia91} The method at fixed
order parameter \cite{Yedidia91} and some of its results are described in 
Refs.\ \cite{Don95,Don91} for the special case of the Hubbard model in
high spatial dimensions ($d\to\infty$). Here we extend these
investigations to {\em all\/} finite dimensions.  As described in Refs.\ 
\cite{Don95,Don91,Yedidia91,Kopietz93} the order parameter $\Delta$
and the N\'eel temperature $T_N$ are determined by the roots of the
optimization equation
\begin{equation}
\label{optim}
\frac{df}{d\Delta} = 0 \; ,
\end{equation}
where $f$ is the free energy density.
It then follows that one has to determine
the order parameter from the selfconsistency condition
\begin{equation}
\label{BCS-selfc}
\Delta = 2h_o \int_0^\infty d\epsilon \; N_d (\epsilon)
\frac{\tanh \left[ \frac{1}{2}\beta \eta (\epsilon) \right]}{\eta (\epsilon)} \; .
\end{equation}
In Eq.\ (\ref{BCS-selfc}) $h_0$ denotes the symmetry-breaking field,
$\eta(\epsilon)=\mbox{sgn}(\epsilon)\sqrt{\epsilon^2 + h_0^2}$ the
dispersion, and $\beta=1/T$ the inverse temperature. 
$N_{d}(y)$ being the density of states in $d$ dimensions.
It is important to
note that in {\em each\/} order $h_0$ and $\Delta$ have to be determined
self-consistently from the Hartree-Fock contribution {\em and\/} the 
fluctuations together to obtain systematic corrections to mean-field theory.

%%%% d=3 %%%%

First we consider sc perturbation theory in the
broken-symmetry phase in dimensions $d\geq 3$.  Fortunately, although
the evaluation of the various higher order diagrams in this approach
at {\em finite\/} values of the interaction ($U>0$) is difficult, if
possible at all, the results at {\em weak\/} coupling (i.e., for $U
\to 0$) are simple: For all $d\geq 3$ one finds that the exact value
of the N\'eel temperature $T_{N}$ and the exact order parameter
$\Delta$, can be expressed in terms of their Hartree equivalents and a
scaling factor $q_{d}$.  As is well-known, the Hartree 
N\'eel temperature is exponentially small for $U\rightarrow 0$:
\[
T_{N}^{H} \sim e^{I_{d}-1/U N_{d}(0)}\; ,
\]
where $I_{d}$ can be expressed in terms of an integral,
\[
I_{d}=\int_{0}^{\infty} dy\,\, \frac{1}{y}\left( \tanh y-1+\frac{N_{d}(y)}
{N_{d}(0)}\right) - \ln 2\; ,
\]

Now, we find that the {\em exact\/} 
expressions for $T_{N}$ and $\Delta (T)$ differ from the mean-field results
by a scaling factor $q_{d}$:
\begin{equation}
\label{TNDelass}
\left. \begin{array}{rcl}
T_{N} &\sim& q_{d}^{\phantom{\dagger}} T_{N}^{H}\\
\Delta (T) &\sim& q_{d}^{\phantom{\dagger}}\Delta^{H}(T/q_{d}^{\phantom{\dagger}})
\end{array} \right. 
  \quad (U \to 0)\; .
\end{equation}
For the special case $d=\infty$ we know from Ref.\ \onlinecite{Don91}
that $q_{\infty} = \exp ( -\gamma )$, where $\gamma = 
\sqrt{2} \mbox{arcoth}(\sqrt{2})$, so that $q_{\infty} \simeq 0.2875$. 
The renormalization factor $q_{d}^{\phantom{\dagger}}$ for all $d\geq 3$ is 
determined by the second order diagrams only and can be conveniently written 
in the form of an exponential of a lattice sum:
\begin{equation}
\label{qdexpl}
q_{d}=e^{-S_d}\; ,
\end{equation}
where $S_{d}$ is given by
\[
S_d= \frac{1}{t} \sum_{|{\bf j}|\;\mbox{\scriptsize{even}}}
      \frac{F_{\bf j}^{2}}{F_{\bf 0}^{3}}
      \int_{0}^{\infty}d\tau\left[ G_{\bf j}(\tau )\right]^{2}
\]
and $F_{\bf j}$ and $G_{\bf j}(\tau )$ are integrals over Bessel functions:
\begin{eqnarray*}
F_{\bf j}&=&\frac{1}{2\pi t}\int_{0}^{\infty}dz\; \prod_{\ell =1}^{d}
          \left[ J_{|j_{\ell}|}(z)\right] \\
G_{\bf j}(\tau )&=&\frac{1}{\pi}\int_{0}^{\infty}dz\;
                 \frac{\tau}{\tau^{2}+z^{2}}\prod_{\ell =1}^{d}
                 \left[ J_{|j_{\ell}|}(z)\right] \; .
\end{eqnarray*}
This expression is formally exact for all $d > 2$. We have
numerically evaluated the renormalization factor $q_{d}$ for
$d=3$ and find the following result:
\[ 
q_{3}=0.2599 \; .
\]
Thus, the exact N\'eel temperature and the exact order parameter
are smaller than the Hartree-Fock predictions by almost a factor
of {\em four\/}.

Several remarks are in order: ($i$) The term labeled by ${\bf j}$ in the
lattice sum $S_{d}$ stems from second order diagrams containing two Hubbard
vertices a distance $|{\bf j}|$ apart. ($ii$) Since we know from previous
calculations in high dimensions\cite{Don91} that sites at a relative
distance $|{\bf j}|$ give a contribution of order $d^{-|{\bf j}|}$ to the
free energy, one expects that the lattice sum $S_{d}$ converges extremely
rapidly if $d$ is large. ($iii$) Keeping only the ${\bf j}={\bf 0}$ term
in $S_{d}$ corresponds to the {\em local approximation\/} of Ref.\ \cite{Niki}.
In $d=3$ we find that $q_{3}^{\mbox{\scriptsize{loc}}}=0.2673$. Comparison of
$q_{3}^{\mbox{\scriptsize{loc}}}\simeq 0.2673$ with the exact result
$q_{3}^{\phantom{\dagger}}\simeq 0.2599$ shows that the local approximation
works very well even in $d=3$. ($iv$) Note that {\em odd\/} lattice sites
(i.e., interactions between the two sublattices) do not contribute to 
$q_{d}^{\phantom{\dagger}}$ to leading order in $U$ for $U \to 0$. 
This is due to a different symmetry of the Green functions for odd and 
even values of $|{\bf j}|$. ($v$) Also note that our result (\ref{qdexpl}) 
for the renormalization factor breaks down for $d=2$ since in this 
case the integrals $F_{\bf j}$ diverge logarithmically for all even $|{\bf j}|$.

%%%% figure %%%%

\begin{figure}[t]
  \epsfxsize7.6cm
  \centering\leavevmode\epsfbox{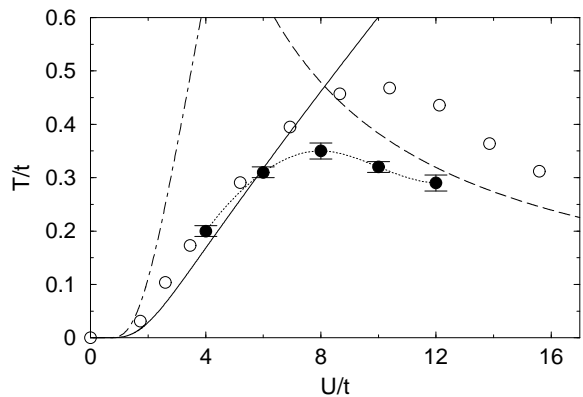}
  \caption{N\'eel temperature $T_N$ as function of $U$ from various
    approaches: Hartree-Fock theory (dot-dashed line), Heisenberg
    limit from high temperature series expansion\protect\cite{Rushbrook74} 
    (dashed line), quantum Monte Carlo simulations in infinite spatial 
    dimensions\protect\cite{Jarrell92} (open
    circles) and three dimensions\protect\cite{Staudt} (filled circles, 
    the dots are meant as a guide to the eye only), and the result of
    this work (solid line).}
\label{figure}
\end{figure}                         

The asymptotic result for the N\'eel temperature in Eq.\ (\ref{TNDelass})
has a relevance beyond the pure weak coupling limit. This is particularly
clear in {\em infinite dimensions\/}, where accurate quantum Monte Carlo data 
are available.\cite{Jarrell92} A comparison of the 
Monte Carlo results with the appropriate asymptotic formula for 
$d=\infty$ \cite{Don91}
reveals excellent agreement for all $U\alt D$. 
Therefore one
expects that (\ref{TNDelass}) yields an equally good approximation 
for $U \alt D$ in $d=3$. The phase diagram of the 
{\em three-dimensional\/} Hubbard model was calulated with a quantum Monte 
Carlo technique by Hirsch\cite{Hirsch87} and Scalettar 
{\it et al.\/}\cite{Scalettar} 
A comparison of the Monte Carlo data from Ref.\ \cite{Scalettar}
with the expected behavior (\ref{TNDelass}) reveals 
a significant discrepancy, suggesting (as was also pointed out in Refs.\
\cite{Kakehashi88,Niki}) that the existing Monte Carlo estimates for $T_{N}$ 
are too high. Improved Monte Carlo simulations\cite{Staudt} show a
significantly reduced N\'eel temperature. Yet it is still more than 15\%
higher than the asymptotic value (see Fig.\ \ref{figure}).
Our weak-coupling approximation formula (\ref{TNDelass}) 
and the strong-coupling approximation formula 
$T_{N}^{\mbox{\scriptsize{sc}}}=3.83 t^{2}/U$ from Ref.\ \cite{Rushbrook74} 
can hence serve as benchmarks for future Monte Carlo simulations.

Since the upper critical dimension of the Hubbard model is presumably $d_{u}=4$, one
expects that the critical behavior of the order parameter in $d=3$ is 
characterized by a nontrivial critical exponent for all $U>0$. Yet the
exact asymptotic behavior of $\Delta (T)$ in (\ref{TNDelass}) displays
mean-field behavior near $T_{N}$, with a critical exponent $\beta =1/2$.
These two observations can easily be reconciled by calculating the size 
$\Delta T$ of the Ginzburg region in $d=3$, which requires as input\cite{NO} 
the correlation length $\xi$ and the jump in the specific heat $\Delta C$ at 
$T_{N}$. An explicit calculation shows that $\xi$ is exponentially large at 
weak coupling, $\xi\propto \left[ T_N ( T_N - T ) \right]^{-1/2}$, while $\Delta C$ 
is exponentially small, $\Delta C\propto T_{N}$. Combination gives for the
relative size of the Ginzburg region: $\Delta T/T_{N}\propto T_{N}^{4}$, which is
exceedingly small at weak-coupling and vanishes for $U \to 0$.

%%%% d=2 %%%%

Now we address sc perturbation theory for the broken-symmetry 
phase in $d=2$. Since $T_N=0$ in $d=2$,\cite{MerminWagner66} we focus on the 
renormalization of the ground state order parameter: $\Delta=q_2\Delta^H$.
Calculating only the second order contribution to Eq.\ (\ref{optim}) we
find $df_2/d\Delta \sim h_0 x^2/6$
in the limit $U \to 0$ with
$x=\frac{U}{4\pi^2t}\left[ \ln \left( \frac{4t}{h_0} 
\right) \right]^2$. The solution of the self-consistency equation
(\ref{BCS-selfc}) shows an {\em interaction-dependent\/}
renormalization:\cite{Kopietz93}
\begin{equation}
\label{d2renorm}
\Delta = \frac{8t}{U} \left( 3 - \sqrt{3} \right)
\exp \left( -\sqrt{ \frac{4 \pi^2 t ( 3 - \sqrt{3} )}{U} } \right) \; .
%q_2 = e^{-\sqrt{\frac{4\pi^2t}{U}}\left( \sqrt{3-\sqrt{3}} -1 \right)}
%(3-\sqrt{3}) \; .
\end{equation}
In contrast to the result in $d \ge 3$, where $q_d$ is constant,
the renormalization factor in $d=2$ vanishes exponentially as $U \to 0$. This
result already indicates that the sc perturbation series in two dimensions
has {\em no} small expansion parameter at weak coupling.\cite{expparam}
Hence higher order fluctuation terms are important to decide whether
the mean-field prediction for the form of the order parameter is at least
{\em qualitatively\/} correct or does totally fail in $d \leq 2$ in the
symmetry-broken phase.

To gain some insight into sc perturbation theory in
higher orders we calculated bubble and ladder diagrams.
We find
\begin{equation}
\label{lad-bub-optim}
\frac{df_{\mbox{\scriptsize BL}}}{d\Delta} \sim h_0 F_{\mbox{\scriptsize BL}}(x)
\end{equation}
with the function 
\begin{eqnarray*}
F_{\mbox{\scriptsize BL}}(x) & = & 
6+x-\frac{x^2}{3}-\frac{1}{2} \ln \left[ (1+x)^{9\left(1+\frac{1}{x}\right)}
     (1-x)^{\left(1-\frac{1}{x}\right)} \right] 
\end{eqnarray*}
for $0 < x < 1$. The important result is that the optimization
condition Eq.\ (\ref{optim}) has {\em no roots\/} for $0 < x < 1$,
apart from the high temperature solution $\Delta = 0$.

We comment on the results. ($i$) Our analysis shows that 
Hartree-Fock theory bears {\em no\/} relevance to the behavior
of the Hubbard model in $d=2$ at half-filling. The mean-field
result is completely destroyed by quantum fluctuations. Hence
the antiferromagnetic order parameter must have a completely different
form: $\ln (4t/h_0) \ll \sqrt{t/U}$.
($ii$) A calculation of bubble and ladder diagrams in $d=\infty$ shows
that all contributions $f_n$ for $n > 2$ yield small
corrections to $q$ of the form $q=q_\infty \exp \left[ 2 N_\infty(0) \gamma^2 U
+ {\cal O}(U^2) \right]$, where $\gamma$ has been defined above.
($iii$) Concerning the behavior
of sc perturbation theory we conclude that only in
dimensions $d > 2$ low-order sc perturbation theory is capable of
describing the physics of the Hubbard model correctly. In low
dimensions sc perturbation theory in the symmetry broken phase
diverges. ($i\nu$) Remarkably, 
in $d=1$ the structure of sc perturbation theory is {\em exactly the
same\/} as in $d=2$, now with
$x=\frac{U}{2\pi t}\ln \left( \frac{2t}{h_0} \right)$ in
Eq.\ (\ref{lad-bub-optim}).

%%%% anisotropic model %%%%

The observation that sc second order perturbation
theory gives even quantitatively exact results in $d=3$ and
diverges in $d=2$ indicates that between two and three spatial
dimensions must be a transition where sc perturbation theory
becomes inadequate. We introduced a weak hopping amplitude
$t_\perp$ in $c$-direction. The anisotropy is then given by the
dimensionless quantity $\lambda = t_\perp / t_\parallel \ll 1$ where
$t_\parallel$ denotes the hopping in the planes. In this
anisotropic three-dimensional model the renormalization factor
can again be expressed as a lattice sum according to Eq.\ (\ref{qdexpl}).
However, the smaller $\lambda$ is the more sites have to be
summed over. In the $j_1,j_2$-plane all points within a region
$1/\lambda^2$ around the origin contribute with $\ln (1 /\lambda )$
to the renormalization factor, yielding
\[
q_\lambda = \exp{\left( -\frac{\mbox{const.}}{\lambda^2 
      \ln (1/\lambda )} \right) } \; .
\]
The renormalization factor is thus strongly reduced in less than
three dimensions and zero in $d=2$. This result demonstrates that
the coupling to the third dimension is essential for
mean-field theories to be qualitatively valid.

%%%% Summary %%%%

In summary, we have investigated the half-filled Hubbard model
in two and three dimensions at weak coupling calculating
systematic corrections to Hartree-Fock mean-field theory. We
have shown that the mean-field antiferromagnetic Hartree-Fock
solution does {\em not\/} yield exact results in the
weak coupling regime. In particular, we showed that quantum
fluctuation induced corrections make sc perturbation theory diverge
in $d \leq 2$. In $d>2$ the exact order parameter and
N\'eel temperture differ from the mean-field result by a scaling factor.

%%%% acknowledgment %%%%

We acknowledge helpful discussions with 
F.\ Gebhard, M.\ Jarrell, A.\ Muramatsu, and T.M.\ Rice.

\widetext


\begin{references}
\bibitem{expparam} From a formal point of view the expansion parameter is
$U \partial f / \partial h_0^2$. This is small of order $U$ in $d > 2$ and of
order unity below.
\end{references}
\end{document}